\documentstyle[epsfig,aps]{revtex}
\begin{document}
\draft
\title{Odd-Spin Yrast States as Multiple Quadrupole-Phonon Excitations} 

\author{N.~Pietralla, P.~von Brentano}
\address{Institut f\"ur Kernphysik, Universit\"at zu K\"oln, 
         50937 K\"oln, Germany}

\author{T.~Otsuka}
\address{Department of Physics, University of Tokyo, Hongo, Bunkyo-ku, 
         Tokyo 113, Japan}
 
\author{R.~F.~Casten}
\address{Brookhaven National Laboratory, Upton, New York 11973, USA}

\date{\today}

\maketitle

\begin{abstract}
The wavefunctions of the lowest odd spin positive parity yrast 
states in the IBA are shown to be nearly pure multiple 
quadrupole-phonon excitations even outside the three dynamical 
symmetries. 
The empirical data for collective nuclei 
with $30\leq Z\leq 80$ confirm 
these predictions. 
The quadrupole-phonon purity of the $2^+_1$ state 
can be measured from E2-branching ratios of the $3^+_1$ 
state. 
These data show a high correlation to the $2^+_1$ Q-phonon 
purity deduced from the E2-decay of $2^+$ 
states.
\end{abstract}

\pacs{21.60.Fw,21.60.Ev,21.10.Re}

Single phonon excitations of atomic nuclei have been well studied 
\cite{BoMo}. 
In the last years multi-phonon modes have also attracted wider 
interest,
e.g.~ \cite{Boe91,Zil91,Cas92,Del93,Kne93,Kor93,Wu93,Zil93,Wu94}. 
Recently the introduction of the quadrupole-phonon (Q-phonon) 
scheme \cite{Siems94,PaduaOts,OtsKimtbp} 
for describing the wavefunctions of low lying collective excitations 
for the O(6)-symmetry of the Interacting Boson Approximation model 
(IBA) \cite{IaAr78,ArIa87} has led to an intuitive understanding of the 
E2-decay of $\gamma$-soft nuclei \cite{Tokyo,Padua}. 
In nature there are no nuclei, however, that exactly exhibit the 
properties of the dynamical symmetries of the IBA. 
Better descriptions of real nuclei are achieved by adding a symmetry 
breaking part to the Hamiltonians of the dynamical symmetries. 
Then, the wavefunctions of the eigenstates have no longer a simple 
structure rendering the understanding of their properties 
more difficult. 
Therefore it is desirable to find simple expressions for the 
wavefunctions of at least some 
nuclear states also {\em outside} 
the analytical limits. 
It will be shown in this Letter that a generalization of the 
Q-phonon scheme can fulfil this 
for the restricted set of the low lying positive 
parity yrast states that we will discuss in the following. 

A pure n-Q-phonon configuration with angular momentum $L$ is obtained 
by an n-fold application of the quadrupole operator to the 
ground state $\mid gs\rangle$
\begin{equation}
\mid L,n\rangle={\cal N}^{(L,n)}\,
(\underbrace{QQ\dots Q}_{n=L/2})^{(L)}\mid gs\rangle
\label{qphe}
\end{equation}
for even $L$ and 
\begin{equation}
\mid L,n\rangle={\cal N}^{(L,n)}\,
(Q(\underbrace{QQ\dots Q}_{n-1=(L+1)/2})^{L+1})^{(L)}
\mid gs\rangle
\label{qpho}
\end{equation}
for odd $L$ where 
$Q=Q^\chi=s^+\tilde{d}+d^+s+\chi(d^+\tilde{d})^{(2)}$ 
is the general quadrupole operator of the IBA which includes a 
$(d^+\tilde{d})$ term and $\cal N$ is a normalization factor. 

For an investigation of the structure of states in the full 
symmetry triangle of the IBA, the properly scaled consistent Q 
(CQF) Hamiltonian
\begin{equation}
H=Q^\chi\cdot Q^\chi-\frac{\epsilon}{\kappa}\,n_d
=H(\epsilon/\kappa,\chi)
\end{equation}
and the E2-transition operator 
\begin{equation}
T(E2)=q\cdot Q^\chi
\end{equation}
are useful \cite{CQF,Lipas85}. 
The limiting cases of the three dynamical symmetries correspond 
to the following parameter combinations: 
U(5): ($\epsilon/\kappa=\infty$, $\chi=0$), 
SU(3): ($\epsilon/\kappa=0$, $\chi=-\sqrt{7}/2$) and for O(6): 
($\epsilon/\kappa=0$, $\chi=0$). 
The parameter $q$ in the E2-transition operator is the effective 
quadrupole charge. 
For arbitrary values of the Hamiltonian parameters the multiple 
Q-phonon configurations are not eigenstates of the Hamiltonian. 
However, they can be expressed as linear combinations of the true 
eigenstates $\mid L_i\rangle$ of the CQF-Hamiltonian $H$
\begin{equation}
\mid L,n\rangle =\sum_i \alpha_i^{(L,n)}\mid L_i\rangle\ .
\label{expa}
\end{equation} 
Here $\mid L_i\rangle$ denotes the $i$-th eigenstate of $H$ 
with total angular momentum $L$ and $\alpha_i^{(L,n)}$ denotes its 
amplitude in the n-Q-phonon configuration. 
If a {\em given} state $L_j$ has a dominant amplitude, i.e. if 
$(\alpha_j^{(L,n)})^2\approx 1$, then this state is in good 
approximation a multi Q-phonon configuration. 

The amplitudes $\alpha_i^{(L,n)}$ in Eq.\ (\ref{expa}) can be 
calculated in the sd-IBA-1 with standard programs like PHINT 
\cite{isa} by the use of the reduced E2 matrix elements 
$r(L_i\rightarrow L_f)$ defined by: 
\begin{equation}
r(L_i\rightarrow L_f)=\frac{\langle L_f\parallel Q\parallel L_i
\rangle}{\sqrt{2L_i+1}}\ .
\label{defr}
\end{equation}
The amplitudes $\alpha_i^{(L,n)}$ can be written as follows: 
\begin{equation}
\alpha_i^{(L,n)}=\frac{(-1)^L\,{\cal N}^{(L,n)}}{\sqrt{2L+1}}\,
\Sigma_{\nu_1\dots\nu_{n-1}}\ r(0_1^+\rightarrow 2_{\nu_1})\,
r(2_{\nu_1}\rightarrow 4_{\nu_2})\dots r([I_{n-1}]_{\nu_{n-1}}
\rightarrow L_i) \ .
\label{alpha}
\end{equation}
This expression is obtained from Eq.\ (\ref{expa}) by multiplying with 
the bra $\langle L_i\mid$ from the left, by decoupling the 
quadrupole operators from Eq.\ (\ref{qphe}) or (\ref{qpho}), and by 
introducing the unity operator 
\mbox{($1=\Sigma_I\Sigma_{\nu(I)} \mid I_\nu\rangle\langle I_\nu\mid$)} 
between every two following quadrupole operators. 
The intermediate spin is $I_{n-1}=2(n-1)=L-2$ for even $L$ and 
$I_{n-1}=2(n-1)=L+1$ for odd $L$. 
These requirements allow unique decoupling of the quadrupole operators. 
The positive constant ${\cal N}^{(L,n)}$ is fixed by the 
normalization condition $\Sigma_i\,{\alpha_i^{(L,n)}}^2=1$. 
It has been shown recently that the lowest members 
of the ground state band with even spin, e.g., the $2^+_1$ and the 
$4^+_1$ states, have a nearly pure Q-phonon configuration 
in the ($\epsilon/\kappa$, $\chi$) space \cite{me}. 

\begin{figure}[hbt]
\centerline{\psfig{figure=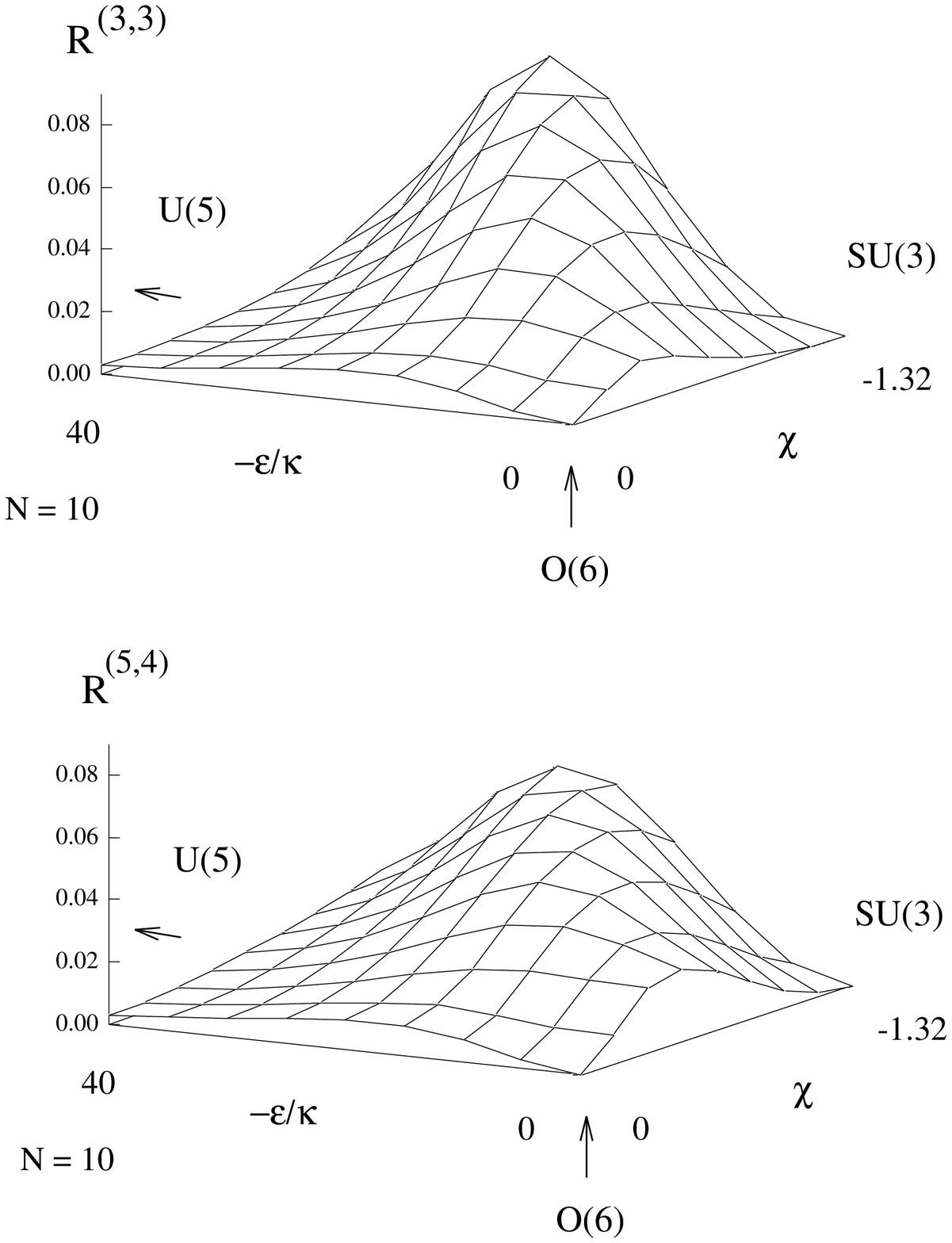,height=20cm,angle=0}}
\caption{
         Q-phonon impurities $R^{(3,3)}=\langle r_3\mid r_3\rangle$ 
         and $R^{(5,4)}=\langle r_5\mid r_5\rangle$ of the $3_1^+$ 
         and the $5^+_1$ states 
         for the whole IBA space calculated gridwise throughout the 
         symmetry triangle for $N=10$ bosons.\label{figr35n10}}
\end{figure}

It is the main purpose of this Letter to show that this is true also 
for the positive parity yrast states with odd spin in the 
full ($\epsilon/\kappa$, $\chi$) parameter space with the exception 
of the SU(3) symmetry. 
A comparison to the exact expressions found in the O(6)-limit 
suggests that the lowest yrast states with odd spin can be written 
as Q-phonon configurations consisting of $n=(L+3)/2$ Q-phonons, where 
$L=3, 5, \dots$. 
In particular one finds: 
\begin{eqnarray}
\mid 3_1^+\rangle =\alpha_3\mid 3,3\rangle +\mid r_3\rangle
=&\alpha_3\,{\cal N}^{(3,3)}(Q(QQ)^{(4)})^{(3)}&\mid 0_1^+\rangle 
+\mid r_3\rangle\ ,\\
\mid 5_1^+\rangle =\alpha_5\mid 5,4\rangle +\mid r_5\rangle
=&\alpha_5\,{\cal N}^{(5,4)}(Q(QQQ)^{(6)})^{(5)}&\mid 0_1^+\rangle 
+\mid r_5\rangle 
\end{eqnarray}
where the labels $\mid L,n\rangle =\mid 3,3\rangle$ and 
$\mid 5,4\rangle$ give the total angular momentum and the 
Q-phonon 
number respectively, as defined in Eq.\ (\ref{qpho}). 
The "rest" wavefunction $\mid r_L\rangle$ is defined by Eqs.\ (8) 
and (9) and the condition that it is orthogonal to the corresponding 
yrast state $\mid L^+_1\rangle$. 
These yrast states have a pure Q-phonon configuration if the rest 
wavefunction vanishes, i.e., if $\langle r_L\mid r_L\rangle =0$ which 
implies $\alpha_3^2=1$ and $\alpha_5^2=1$ (see Eqs.\ (8) and (9)) 
as is the case in the O(6)-symmetry. 
Note that we use "state" for the eigenstates of the Hamiltonian as 
opposed to "Q-phonon configuration" defined in Eqs.\ (\ref{qphe}) 
and (\ref{qpho}). 
Therefore the amount of the {\em Q-phonon impurity} of an eigenstate 
of $H$ is expressed by the norm of the rest wavefunction: 
$R^{(3,3)}=\langle r_3\mid r_3\rangle=1-\alpha_3^2$ and 
$R^{(5,4)}=\langle r_5\mid r_5\rangle=1-\alpha_5^2$. 
For the boson number $N=10$ the Q-phonon impurities 
$R^{(3,3)}$ and $R^{(5,4)}$ 
are shown in Fig.\ \ref{figr35n10} for the 
CQF Hamiltonian as a function of the parameters $\chi$ and 
$\epsilon/\kappa$. 
It is concluded that at least the lowest odd spin positive parity 
yrast states have a rather pure Q-phonon configuration. 
One finds that the squared amplitudes of the rest wavefunctions 
$\langle r_3\mid r_3\rangle$ and $\langle r_5\mid r_5\rangle$ 
are less than 10\%. 

The maximum of the Q-phonon impurity of the $3^+_1$ state is 8\% 
for $N=10$ and 5\% for $N=6$ and for the $5^+_1$ state it is 6\% 
for $N=10$ and  3\% for $N=6$. 
The maxima of the phonon impurities are reached for parameter 
combinations ($\chi=-\sqrt{7}/2$,\ $\epsilon/\kappa\neq 0$) that 
do not correspond to real nuclei. 
Even well deformed nuclei such as $^{168}$Er or $^{178}$Hf are 
described best with values of the parameter $\mid\chi\mid<0.7$ 
\cite{CaWa88}. 
For real nuclei the Q-phonon impurity of these states is 
therefore predicted to be considerably less than 10\%. 
Furthermore, the Q-phonon impurity decreases with increasing spin 
and decreasing boson number. 
Here it should be noted that the Q-phonon picture of odd spin 
states is not defined in the SU(3)-symmetry because, there, no 
E2-transitions can lead out of the ground state band which consists 
only of states with even spin because $Q^{\chi=-\sqrt{7}/2}$ is a 
generator of the subgroup SU(3) and cannot connect different 
irreducible representations of SU(3). 
Infinitesimally breaking of the SU(3)-symmetry allows one to calculate 
the Q-phonon purity. 
The $3^+_1$ and $5^+_1$ states then have practically pure 
Q-phonon configurations as in the two other symmetries. 
But the normalization factor ${\cal N}$ goes to 
infinity when the SU(3) singularity is approached. 
However, as mentioned above, for describing real nuclei a $\chi$-value 
is needed which considerably deviates from its value in the 
SU(3)-symmetry. 
So, problems with the definition of the Q-phonon purity for the 
odd spin states in the 
SU(3)-symmetry do not affect the conclusions that are drawn in 
this Letter concerning atomic nuclei.
 
Of course one could also consider the $3^+_1$ state as a 2-Q-phonon 
configuration ${\cal N}^{(3,2)} (QQ)^{(3)}\mid 0^+_1\rangle$. 
This suggestion fails, however, in all three dynamical symmetries 
of the IBA whereas the configuration 
${\cal N}^{(3,3)} (QQQ)^{(3)}\mid 0^+_1\rangle$ fails only in the 
SU(3)-limit.  
The antisymmetric coupling of the two quadrupole 
operators $(Q^\chi Q^\chi)^{(3)}=(1-4\chi^2/7) (d^+
\tilde{d})^{(3)}$ vanishes identically in the SU(3) dynamical 
symmetry ($\chi=-\sqrt{7}/2$). 
In addition it is a generator of the O(5) Lie-subgroup and 
must conserve the $d$-boson seniority $\tau$. 
Therefore, it cannot connect the ground state ($\tau=0$) 
to any $3^+$ state ($\tau=3$) if the seniority is a good 
quantum number, as it is on the whole transition between 
the U(5) and O(6) dynamical symmetries.  
We note that this problem and problems of uniqueness 
of configurations can be completely solved for the O(6)-symmetry 
by ref.~\cite{OtsKimtbp} by application of a proper symmetrizer 
to the Q-phonon operators. 
The configurations created by Eq.\ (\ref{qphe}) are identical to 
those symmetrized. 
Also Eq.\ (\ref{qpho}) produces Q-phonon configurations 
which in practice coincide with the symmetric ones. 

\begin{figure}[hbt]
\centerline{\psfig{figure=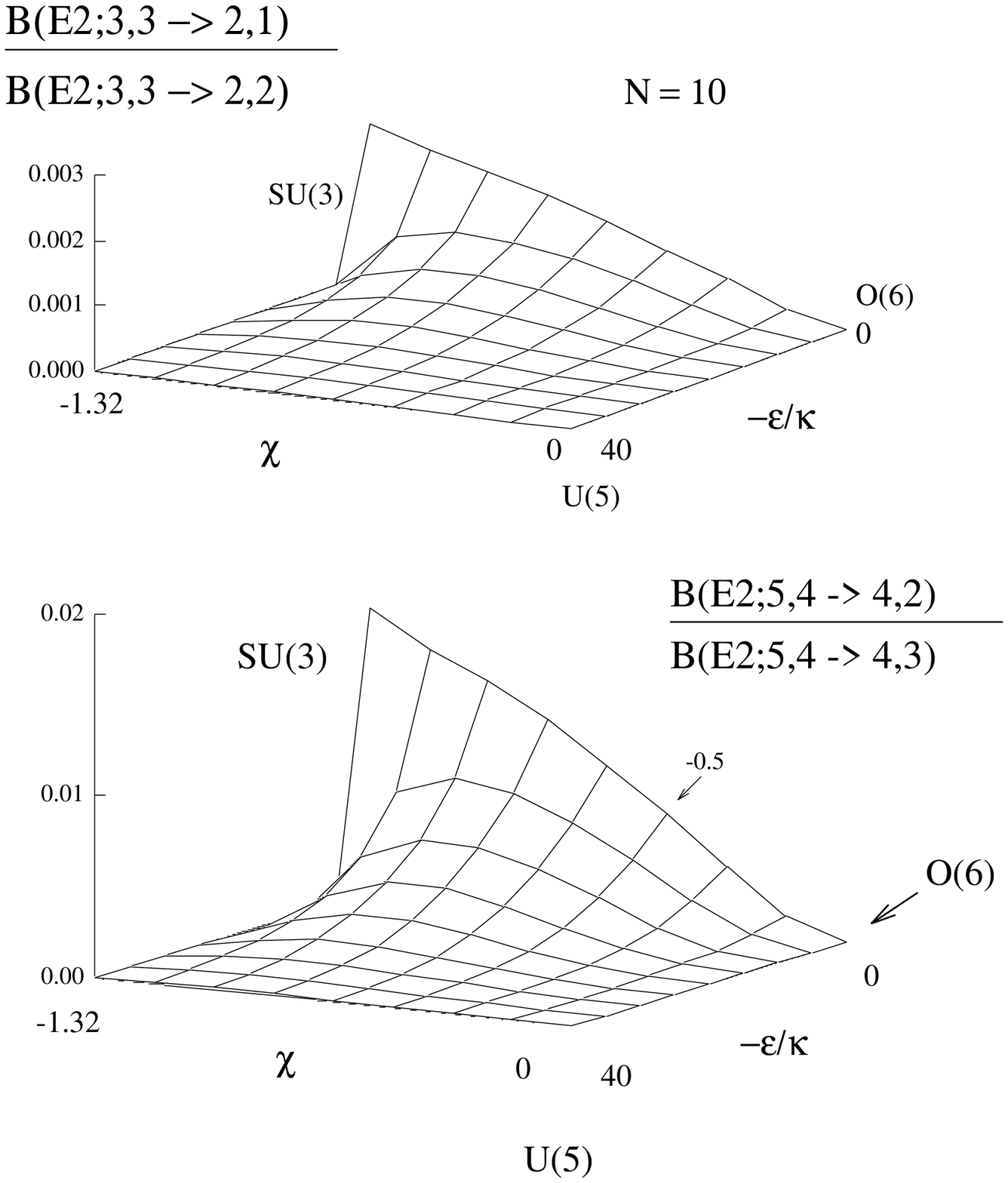,height=20cm,angle=0}}
\caption{
         E2-branching ratios for pure Q-phonon configurations 
         as defined by Eqs.\ (10) and (11) 
         for the whole IBA structure space calculated gridwise 
         for $N=10$ bosons. 
         Transitions between configurations that differ by more 
         than one Q-phonon are hindered by 2 orders of magnitude. 
         The selection rule is better fulfilled for the lower 
         spin. 
         Note that the orientation of the plot differs from that in 
         Fig.~1 for ease of visualization.  \label{figphobra}}
\end{figure}

Combining the present results for the odd spin states with the 
previous results for the even spin yrast states \cite{me} one finds 
that the low spin yrast states can be 
considered approximately pure Q-phonon configurations. 
This applies even for deformed nuclei outside the SU(3)-symmetry. 
One additional advantage of this description consists in the 
occurence of approximate selection rules for electromagnetic 
E2-decay between {\em pure} Q-phonon configurations with different 
numbers of Q-phonons. In order to demonstrate the selection rules we 
have investigated the E2-decay properties of the pure Q-phonon 
configurations of Q-phonon number of $n=3,4$ and total angular 
momenta $L=3,5$ respectively as defined in Eqs.\ (\ref{qpho}), 
(\ref{expa}), and (\ref{alpha}). 
In Fig.\ \ref{figphobra} the 
ratios of the E2-transition strengths 
$B(E2;\mid L,n\rangle\rightarrow\mid L^\prime,n^\prime\rangle )$ 
between pure Q-phonon configurations are shown: 
\begin{equation}
\frac{B(E2;\mid 3,3\rangle \rightarrow \mid 2,1\rangle)}
     {B(E2;\mid 3,3\rangle \rightarrow \mid 2,2\rangle)} 
\end{equation}
and 
\begin{equation}
\frac{B(E2;\mid 5,4\rangle \rightarrow \mid 4,2\rangle)}
     {B(E2;\mid 5,4\rangle \rightarrow \mid 4,3\rangle)} 
\end{equation}
The final configurations in the denominators are coupled as follows: 
$\mid 2,2\rangle = {\cal N} (QQ)^{(2)}\mid 0^+_1\rangle$ and 
$\mid 4,3\rangle = {\cal N} (Q(QQ)^{(4)})^{(4)}\mid 0^+_1\rangle$ 
respectively. 
From these calculations one finds that these branching ratios are 
very small in the full symmetry triangle. 
Therefore rather good selection rules exist for E2-transitions 
between pure Q-phonon configurations. 

\begin{figure}[hbt]
\centerline{\psfig{figure=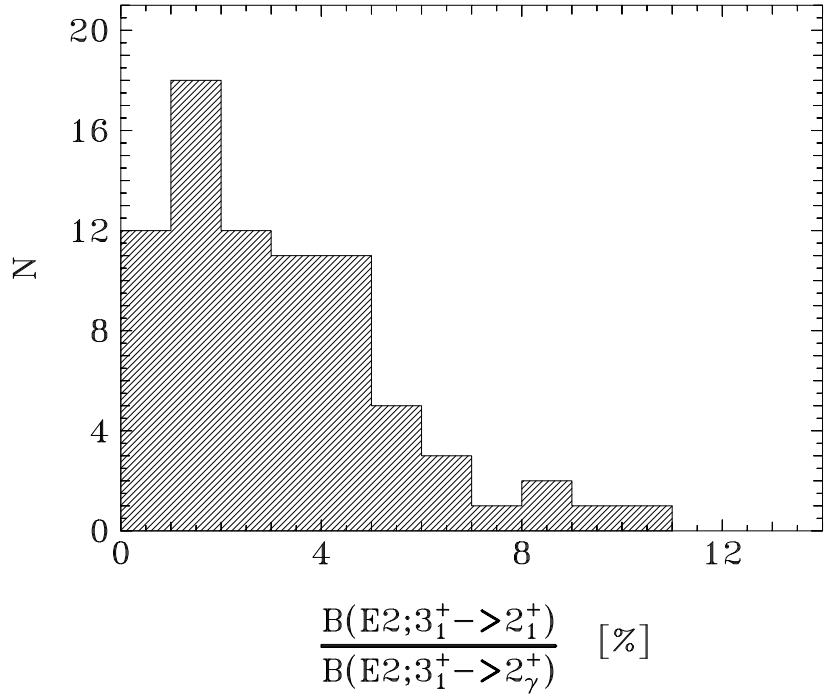,height=20cm,angle=0}}
\caption{
         Histogram of all experimentally known branching ratios 
         $[$23$]$ from the $3^+_1$ to the lower $2^+$ states 
         (see text) for collective nuclei 
         ($E_{4^+_1}/E_{2^+_1}>1.9$) with $30\leq Z\leq 80$. 
         In all cases pure E2-transitions were assumed.  
         \label{fighist}}
\end{figure}

This observation leads to simple predictions for experimental 
E2-branching ratios. 
From recent work \cite{me} it is known that, empirically, 
the one Q-phonon configuration 
\begin{equation}
\mid 2,1\rangle={\cal N}^{(2,1)}Q\mid 0^+_1\rangle
\approx\alpha\mid 2^+_1\rangle +\gamma\mid 2^
+_\gamma\rangle;\hspace{0.1cm} \gamma^2\leq 7\%
\end{equation}
consists mainly of the first excited state with a 
small admixture of the $2^+_\gamma$ state. 
We use the notation $2^+_\gamma$ for either the $2^+_2$ state or 
the $2^+_3$ state, whichever is connected to the $0^+_2$ state by 
only a small E2 matrix element. 
The first excited state has therefore the form 
\begin{equation}
\mid 2^+_1\rangle\approx\frac{1}{\alpha}\mid 2,1\rangle 
+\epsilon\mid 2^+_\gamma\rangle; 
\hspace{0.3cm}\epsilon^2=(\frac{\gamma}{\alpha})^2\leq 8\% \ .
\end{equation}
The branching ratio for E2-transitions of the $3^+_1$ state to 
the $2^+_1$ and the $2^+_\gamma$ states is given by 
\begin{equation}
\frac{B(E2;3^+_1\rightarrow 2^+_1)}{B(E2;3^+_1\rightarrow 
2^+_\gamma)}\approx 
\left(\frac{\langle 3,3 \parallel Q\parallel(\frac{1}{\alpha}
\mid 2,1\rangle +\epsilon\mid 2^+_\gamma\rangle)}
{\langle 3,3\parallel Q\parallel 2^+_\gamma 
\rangle}\right)^2\approx\epsilon^2\approx 1-\alpha^2 \ . 
\label{impbra3}
\end{equation}
This E2-branching ratio from the $3^+_1$ to 
the $2^+_1$ and the $2^+_\gamma$ states is thus a measure of the  
Q-phonon impurity $1-\alpha^2$ of the $2^+_1$ state. 
An alternative measure of this Q-phonon impurity of the $2^+_1$ state 
is the ratio 
\begin{equation}
R^{(2,1)}=\frac{\sum_{i>1}B(E2;0^+_1\rightarrow 2^+_i)}
{\sum_{i\geq 1}B(E2;0^+_1\rightarrow 2^+_i)}=1-\alpha^2
\label{r21}
\end{equation}
which has been studied in ref.\ \cite{me}. 
Fig.\ \ref{fighist} shows according to Eq.~(\ref{impbra3}) the 
experimentally known branching ratios 
for the $3^+_1$ state for collective nuclei 
($E_{4^+_1}/E_{2^+_1}>1.9$) with $A<200$ \cite{NDS}. 
This histogram confirms the parameter-free prediction of the 
sd-CQF-IBA-1 \cite{me} that the Q-phonon purity of the $2^+_1$ 
state exceeds  93\% for all nuclei in the symmetry triangle. 
The largest deviations from a pure Q-phonon configuration are found 
for the $2^+_1$ states of the Os isotopes which are transitional 
nuclei and lie between the O(6) and SU(3) symmetry. 
Interestingly, it is also for the O(6) $\rightarrow$ SU(3) transition 
that the IBA predicts 
the largest Q-phonon impurities of the $2^+_1$ state \cite{me}. 

\begin{figure}[hbt]
\centerline{\psfig{figure=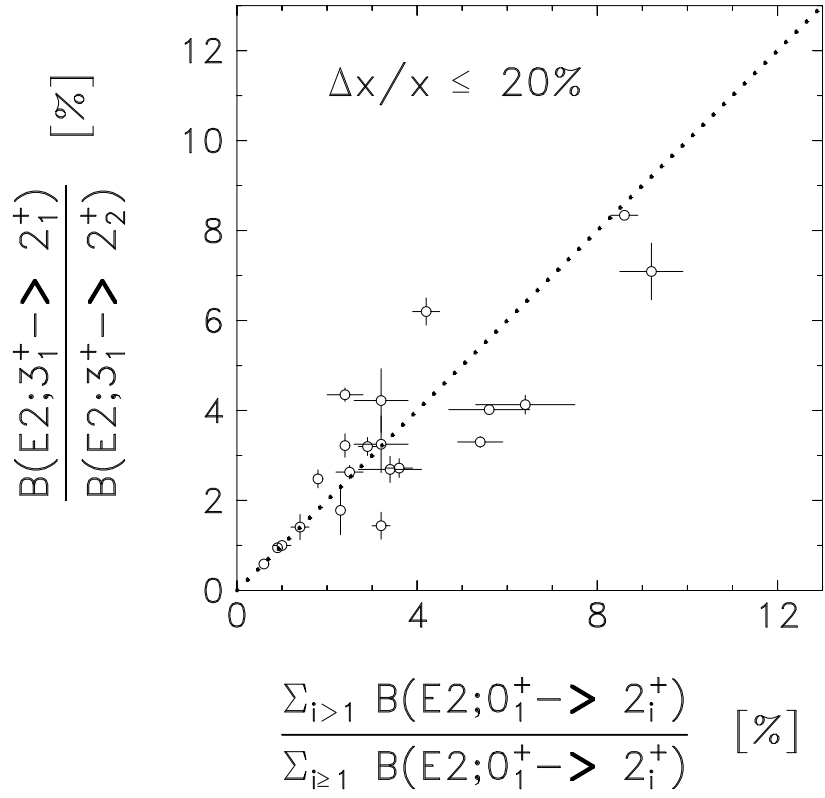,height=20cm,angle=0}}
\caption{
         Correlation of the two "purity" observables 
         for those collective nuclei with $30\leq Z\leq 80$ 
         for which both observables are known with an 
         experimental error of less than 20\%. 
         The dotted line shows a perfect correlation.  
         All data have been taken from $[$23$]$. 
         \label{figkorr}}
\end{figure}

The two E2-branching ratios of Eqs.\ (\ref{impbra3}) and (\ref{r21}) 
involve different observables and thus 
are independent measures of the purity of the Q-phonon 
configuration of the $2^+_1$ state. 
Therefore, the Q-phonon purity of the $2^+_1$ state obtained with the 
two methods can be compared. 
The empirical correlation of these two "purity" observables is 
plotted in Fig.\ \ref{figkorr} for those nuclei for which both 
observables are known experimentally with a relative error smaller 
than 20\%. 
The good correlation is clear; in fact, the correlation coefficient 
of $r\approx 0.84$ is rather large. 
These results confirm the predictive strength of the 
Q-phonon picture even outside the three dynamical symmetries. 

To conclude, we have shown that the lowest odd spin yrast 
states have nearly pure Q-phonon configurations in the IBA. 
Their Q-phonon purity is always larger than 90\% and is calculated 
to increase with increasing spin and decreasing boson number. 
Between pure Q-phonon configurations there exist rather good 
selection rules for E2-transitions. 
Examples of these selection rules are $QQ\not \rightarrow 0$, 
$QQQ\not\rightarrow Q$, or $QQQQ\not\rightarrow QQ$. 
These approximate selection rules improve with 
decreasing spin and increasing boson number. 
The E2-branching ratio of the $3^+_1$ to the $2^+_1$ and the 
$2^+_\gamma$ states is predicted by the CQF-IBA to be less than 
8\% in the full ($\epsilon/\kappa$, $\chi$) parameter space 
(except at the SU(3) dynamical symmetry where the Q-phonon is not defined) 
and it correlates to the Q-phonon purity observable $R^{(2,1)}$. 
These predictions are confirmed by the experimental data. 
The present Q-phonon picture is similar to the ordinary phonon 
concept for the vibrational nuclei but its range of application 
extends to the more deformed nuclei. 
Independent of the actual nuclear shape the Q-phonon scheme provides 
an intuitive understanding of low lying collective 
{\em yrast spin} excitations of nuclei. 
The study of non-yrast states is beyond the scope of the present paper. 
Following ref.~\cite{OtsKimtbp} the phonon picture has to be extended 
to include, e.g., symmetrization. 
This is considerably more complicated than the present approach, which 
works only for yrast states. 
Furthermore, literal extensions do not make the members of the 
(quasi-)$\gamma$-band or the (quasi-)$\beta$-band describable 
in terms of pure Q-phonon configurations, e.g., the $2^+_\gamma$ as an 
approximately pure $(QQ)^{(2)}\mid 0^+_1\rangle$ state. 
This is an interesting limitation of the present approach for 
non-yrast states, which we are presently investigating. 

In the past, most IBA calculations have dealt with fits to a 
specific nucleus, to specific groups of nuclei (e.g., isotopic 
sequences), or to properties of specific classes of nuclei 
(e.g., axially deformed or $\gamma$-soft). 
With refs. \cite{me,CasBre94} and the present work, 
studies of the IBA have entered a new phase focusing on its 
predictions of very general properties of collectivity in nearly 
{\em all} nuclei, essentially {\em independent} of parameters. 
The fact that, in every case studied, these properties are 
confirmed experimentally shows that the truncation embodied 
in the IBA, especially in the CQF, reflects nearly universal 
manifestations of nuclear collectivity. 
This merits microscopic study to better understand the underlying 
origins of collectivity in nuclei. 

For fruitful discussions we thank 
A.~Gelberg, R.~V.~Jolos, O.~Vogel and A.~Zilges. 
This work has been partly supported by the 
{\em Deutsche Forschungsgemeinschaft} under contract 
no.~Br 799/5-2, by the cooperation agreement of the 
{\em Deutsche Forschungsgemeinschaft} and the 
{\em Japan Society for the Promotion of Sciences}, by Grant-in-Aid 
for Scientific Research on International Scientific Research 
Program (No.~0604\,4249) and that for General Scientific Research 
(No.~0680\,4013) from MESC in Japan, and by the 
USDOE under contract DE-AC02-76CH00016.

\end{document}